\documentclass[%
 reprint,
 amsmath,amssymb,
 aps,
]{revtex4-1}

\usepackage{graphicx}
\usepackage{dcolumn}
\usepackage{bm}

\begin{document}

\preprint{APS/123-QED}

\title{Melting of scalar hadrons in an AdS/QCD model modified by a thermal dilaton}

\author{Alfredo Vega}
\email{alfredo.vega@uv.cl}

\author{M. A. Martin Contreras}%
\email{miguelangel.martin@uv.cl}
\affiliation{%
 Instituto de F\'isica y Astronom\'ia, \\
 Universidad de Valpara\'iso,\\
 A. Gran Breta\~na 1111, Valpara\'iso, Chile
}%

\date{\today}

\begin{abstract}


We consider an AdS/QCD model at finite temperature with a dilaton field that we call thermal because, in addition to depending on the holographic coordinate, it also depends on temperature. We study two thermal dilatons in this work such that at $T=0$ they are reduced to the dilatons used by some authors. With these thermal dilatons it is possible to obtain melting temperatures for mesons close to $180$ MeV and also make predictions for other hadrons. We use a procedure based on the analysis  of the holographic potential related to the E.O.M. for modes dual to hadrons to calculate melting temperatures. This technique is easy to implement; therefore, it could be considered a first fast testing criterion for  soft wall-like models at finite temperature.

\begin{description}
\item[PACS numbers]
11.25.Tq, 12.38.Mh, 12.39.Mk
\end{description}
\end{abstract}

\pacs{Valid PACS appear here}
\maketitle


\section{Introduction}


The understanding of how temperature or a very dense medium can affect hadronic properties is a topic that has  attracted the interest of several physicists, who in their attempt to clarify the hadron phenomenology under these extreme conditions,  have built research facilities to do a lot of complex experiments (e.g., see \cite{Adcox:2004mh,Gyulassy:2004zy, Shuryak:2004cy, Song:2012ua}). They have also developed theoretical tools such as lattice QCD (e.g., see \cite{Asakawa:2013laa, Asakawa:2003re, Nakahara:1999vy}), sum rules (e.g., see \cite{Ayala:2016vnt}) or potential models (e.g., see \cite{Mocsy:2008eg, Shi:2013rga}) to approach hadron phenomenology in these extreme conditions. Together with these achievements, throughout the last 20 years, a set of techniques has been added based on gauge/gravity dualities (e.g., see \cite{Maldacena:1997re, Witten:1998qj, Witten:1998zw, Gubser:1998bc, Fujita:2009ca, Fujita:2009wc, Braga:2016wkm}), where temperature and medium effects are introduced by considering black hole metrics in the model formulation.


AdS/QCD models allow the calculation of spectral functions (e.g., see \cite{Fujita:2009wc, Fujita:2009ca, Braga:2016wkm, Colangelo:2013ila, Colangelo:2012jy, Braga:2017bml}) in order to  obtain hadronic properties such as the mass spectrum, decay widths and melting temperatures for different hadrons in the thermal bath. The last property could also be obtained from a holographic potential related to the AdS modes dual to hadrons in the background \cite{Fujita:2009ca, Miranda:2009qp, Bellantuono:2014lra, Bellantuono:2014lra, Vega:2017dbt, Ibanez:2018xci}. In fact, this is possible because the potential has a depth that decreases when the temperature rises until it disappears. When the depth finally disappears at some specific temperature, we say that the hadrons have melted down in the thermal bath. This disappearance temperature is what we call the melting temperature for the hadron.  


In the literature, it is possible to find two alternatives to obtain the holographic potential mentioned above. In both cases, melting temperatures are close to each other, and also with the results calculated from the spectral function. For this reason, this technique, in the simplest case, could be useful for obtaining good (and fast) estimations for melting temperatures \cite{Vega:2017dbt, Ibanez:2018xci}. One alternative considers the Liouville transform \cite{Miranda:2009qp, Bellantuono:2014lra, Bellantuono:2014lra, Vega:2017dbt, Ibanez:2018xci} to transform the equation of motion for hadron dual modes on an AdS -- BH background with dilaton into a Schr\"odinger--like equation, allowing the extraction of the holographic potential (written in terms of the Regge-Wheeler tortoise coordinate)  for later analysis. The other approach is easier, and it is based on the Bogolyubov transform \cite{Fujita:2009ca}. In this case, the holographic potential depends on the hadron mass considered. But using its value at zero temperature, the melting temperature obtained is close to the one calculated using the Schrodinger--like potential or the spectral function techniques.\cite{Vega:2017dbt, Ibanez:2018xci}.


The simplicity when obtaining the holographic potential by means of the Bogolyubov transformation for different AdS -- BH metrics and dilatons and its subsequent uses to calculate melting temperatures make this procedure a useful analytical tool: the method allows us to do an initial, quick test for AdS/QCD models at finite temperature. This compels us to consider two different AdS/QCD models.


The dilatons considered here are modifications of the quadratic dilaton used regularly in AdS/QCD approaches to calculate several hadronic properties such as masses (e.g., see \cite{Vega:2008af, Branz:2010ub, Gutsche:2011vb, Gutsche:2017oro, Dosch:2016zdv, Dosch:2015bca, Dosch:2015nwa}), form factors (e.g., see \cite{Brodsky:2007hb, Brodsky:2011xx, Abidin:2009aj, Abidin:2009hr, Gutsche:2012bp, Gutsche:2012wb, Gutsche:2015xva, Gutsche:2017lyu}), GPDs \cite{Vega:2010ns}, etc. Although these models with the quadratic dilaton have been successful in several uses, it is well known that they must be improved at $T=0$ and at finite temperature also. It is worth noting that at zero temperature the usual soft wall model is not able to realize chiral symmetry breaking \cite{Karch:2006pv, Colangelo:2008us}, and at $T \neq 0$, the  melting temperatures for mesons in a thermal bath are very low \cite{Colangelo:2009ra, Vega:2017dbt}.


Although the dilatons we use are not dynamical, the existing relations between metrics and dilatons suggest to us a sort of phenomenological dilaton, which we call thermal dilaton, because it depends on both the holographic coordinate and the temperature. Thus, we will consider a thermal version of the quadratic dilaton. Moreover, a dilaton that interpolates between two quadratic slopes at low and high $z$ \cite{Gherghetta:2009ac} makes it possible to incorporate chiral symmetry breaking in AdS / QCD models in a satisfactory way at $T=0$. As we will see, in both cases the melting temperatures obtained are low when we consider the traditional case (not thermal). But we can improve this situation in a simple way by extending the dilaton to its thermal version.


This paper consists of four sections. Apart from the introduction, in section II we discuss how to obtain the holographic potential, which will be analyzed to obtain the melting temperatures. In section III we develop the thermal dilaton idea by suggesting two possible forms of temperature dependent dilatons. At the end, in section IV, we discuss results and give conclusions about this work.

\begin{center}
\begin{figure*}
  \begin{tabular}{c c c}
    \includegraphics[width=3.4 in]{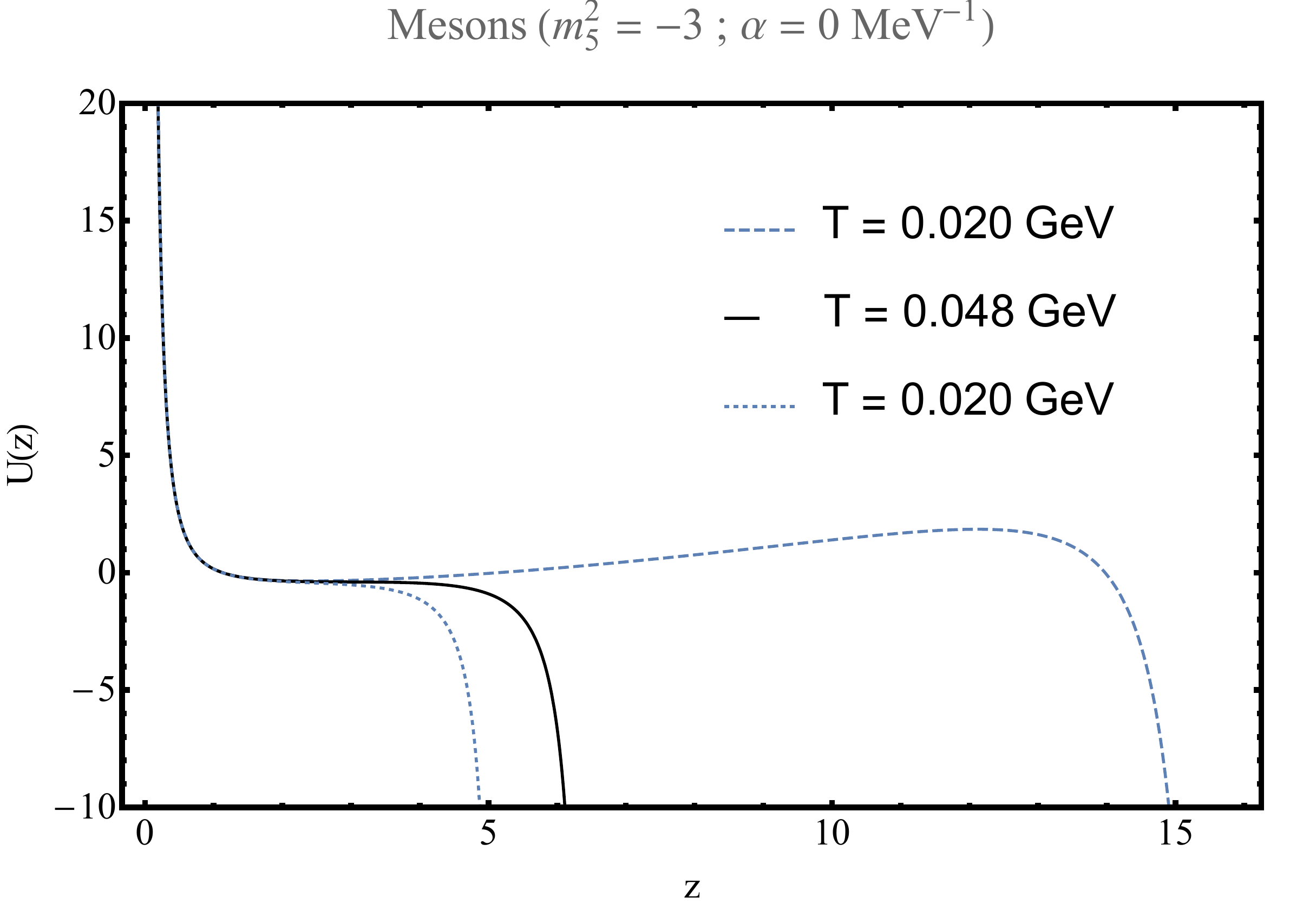}
    \includegraphics[width=3.4 in]{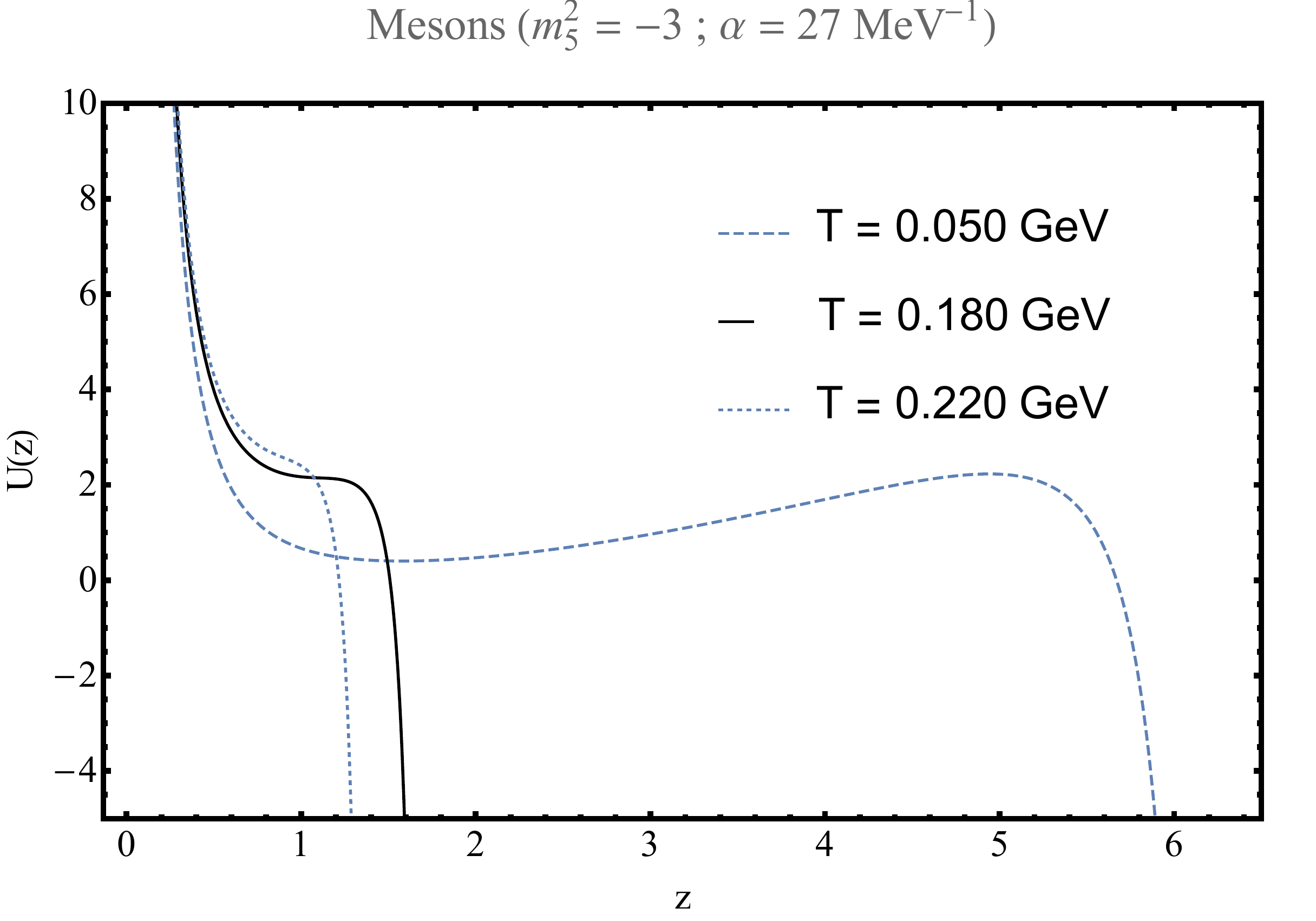}
  \end{tabular}
\caption{
These plots show the holographic potential calculated by using (\ref{potencial}) considering a quadratic thermal dilaton. On the right side, the traditional case ($\alpha = 0$) is considered, where the continuous line is calculated with the prediction of the model for melting temperature for this case, which is 48 MeV. On the left side, we plot potentials considering a thermal quadratic dilaton with $\alpha = 27$ MeV$^{-1}$ that was fitted to obtain 180 MeV for mesonic melting temperature.}
\end{figure*}
\end{center}




\section{Model}

Here we are going to restrict us to the scalar case and we will use the conventions followed in \cite{Vega:2017dbt}.We start from the action for the scalar field in a curved 5D space with the static dilaton given by

\begin{equation}
S_=\frac{1}{2\,K}\int d^{5}x \,\sqrt{-g}\,e^{-\phi \left( z\right) } \mathcal{L},
\end{equation}
where
\begin{equation}
\mathcal{L} = 
g^{MN}\,\partial _{M} \psi \left(x,z\right)\, \partial _{N} \psi \left( x,z\right)
+m_{5}^{2} \psi^{2}\left( x,z\right).
\end{equation}

The metric considered is 
\begin{equation}
ds^{2}=e^{2\,A\left( z\right) }\left[ -f\left( z\right)
dt^{2}+\sum_{i=1}^{3}\left( dx^{i}\right) ^{2}+\frac{1}{f\left( z\right) }\,
dz^{2}\right],
\end{equation}
or
\begin{equation}
g_{MN}=e^{2\,A\left( z\right) }\text{ diag}\left( -f\left( z\right) ,\,1,\,1,\,1,%
\,\frac{1}{f\left( z\right) }\right).
\end{equation}

The E.O.M. associated with this action is
\begin{widetext}

\begin{equation}
e^{B\left( z\right) }\,f\left( z\right) \,\partial _{z}\left[ e^{-B\left(
z\right) }\,f\left( z\right) \,\partial _{z}\psi \right] -f\left( z\right)\,
e^{2A\left( z\right) }\,m_{5}^{2}\,\psi +\omega ^{2}\,\psi -f\left( z\right)\,
q^{2}\,\psi =0,
\end{equation}
where $B(z) = \phi(z) - 3A(z)$.

Considering  our particles at rest, we fix ($\vec{q} = \vec{0}$). Thus the previous equation looks like

\begin{equation}
\label{EOM}
\partial _{z}\left[ e^{-B\left( z\right) }\,f\left( z\right) \,\partial _{z}\psi %
\right] +\left[ \frac{\omega ^{2}}{e^{B\left( z\right) }\,f\left( z\right) }
-e^{-\phi \left( z\right) +5A\left( z\right) }\,m_{5}^{2}\right] \psi =0,
%
\end{equation}
\end{widetext}
where $\omega^{2}$ is related to the hadron mass in the thermal bath. Starting from the last equation, it is possible to obtain the spectral function, and with this, it is possible to study the effect of the temperature on the hadron masses and find the temperature for the different species melted in this medium.


On the other hand, this equation can be transformed into a Schr\"odinger-type equation, with an associated holographic potential. Analyzing this potential it is possible to make estimates of the melting temperature \cite{Fujita:2009ca, Miranda:2009qp, Bellantuono:2014lra, Bellantuono:2014lra, Vega:2017dbt, Ibanez:2018xci}. The holographic potential mentioned above is obtained by means of a Liouville transformation, as in \cite{Vega:2017dbt, Ibanez:2018xci}. Another possibility is to use a Bogoliubov transform \cite{Fujita:2009ca}, which is easier to implement. In the latter case we need to know the hadron mass, but using its value at $T=0$, it is possible to obtain values closer to those obtained by considering the first transformation or working directly with spectral function poles. In our opinion, this is the advantage this procedure. If we could infer the hadron masses, this method is a good alternative for fast and easy first trials of many different ingredients in AdS/QCD models at finite temperature. 

With respect to $m_{5}$, if we keep on mind the AdS/QCD dictionary, it is related to the dimension $\Delta$ of different operators that create hadrons. The expression that relates both quantities is given by

\begin{equation}
m^2_5\, R^{2} = \Delta (\Delta - 4),
\end{equation}
where $R$ is the AdS radius, that can be fixed to be 1. With this expression, we can consider different scalar hadrons in our model, e.g. mesons ($\Delta=3;\,m^{2}_{5}=-3$), glueballs ($\Delta=4;~m^{2}_{5}=0$), hybrid mesons ($\Delta=5;~m^{2}_{5}=5$), tetraquarks ($\Delta=6;~m^{2}_{5}=12$), etc.

Since $u(z) = \sqrt{e^{-B(z)} \, f(z)} \, \psi (z)$ in (\ref{EOM}), it is possible to obtain an equation like $-u''(z) + U(z)\, u(z) = 0$ \cite{Fujita:2009ca}, where $U(z)$ is the thermal holographic potential given by

\begin{widetext}
\begin{equation}
\label{potencial}
U(z) = \frac{e^{2 \,A(z)}}{f(z)}\, m^{2}_{5} - \frac{B'(z)\, f'(z)}{2 \,f(z)} + \frac{f''(z)}{2\, f(z)} - \frac{f'(z)^{2}}{4\, f(z)^{2}} - \frac{\omega^{2}}{f(z)^{2}} + \frac{B'(z)^{2}}{4} - \frac{B''(z)}{2}.
\end{equation}
\end{widetext}

From this holographic potential we will  obtain the melting point of some scalar hadrons by studying the temperature at which the well disappears. In this case, we use the hadron mass calculated at $T=0$.




\begin{center}
\begin{figure*}
  \begin{tabular}{c c c}
    \includegraphics[width=3.4 in]{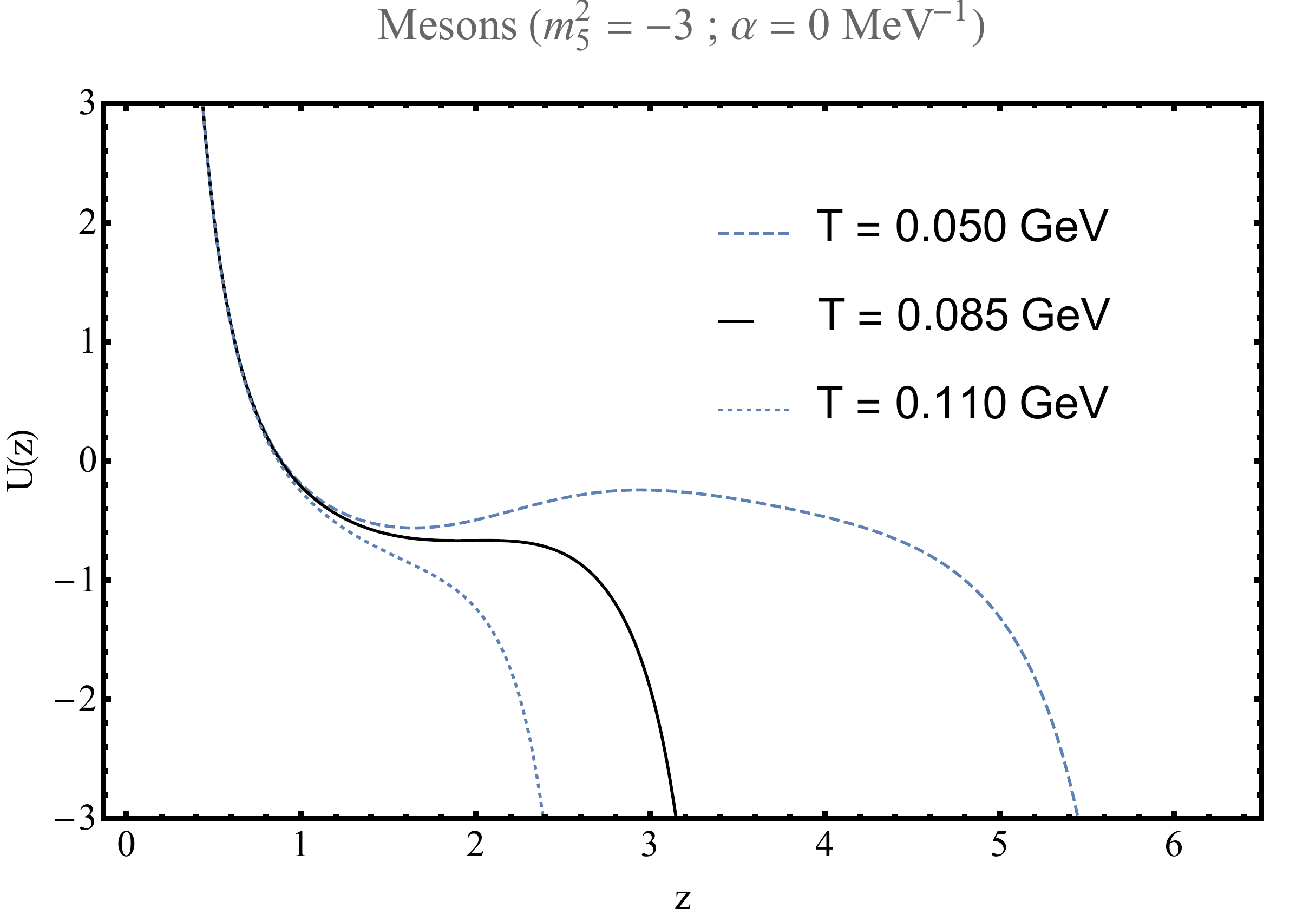}
    \includegraphics[width=3.4 in]{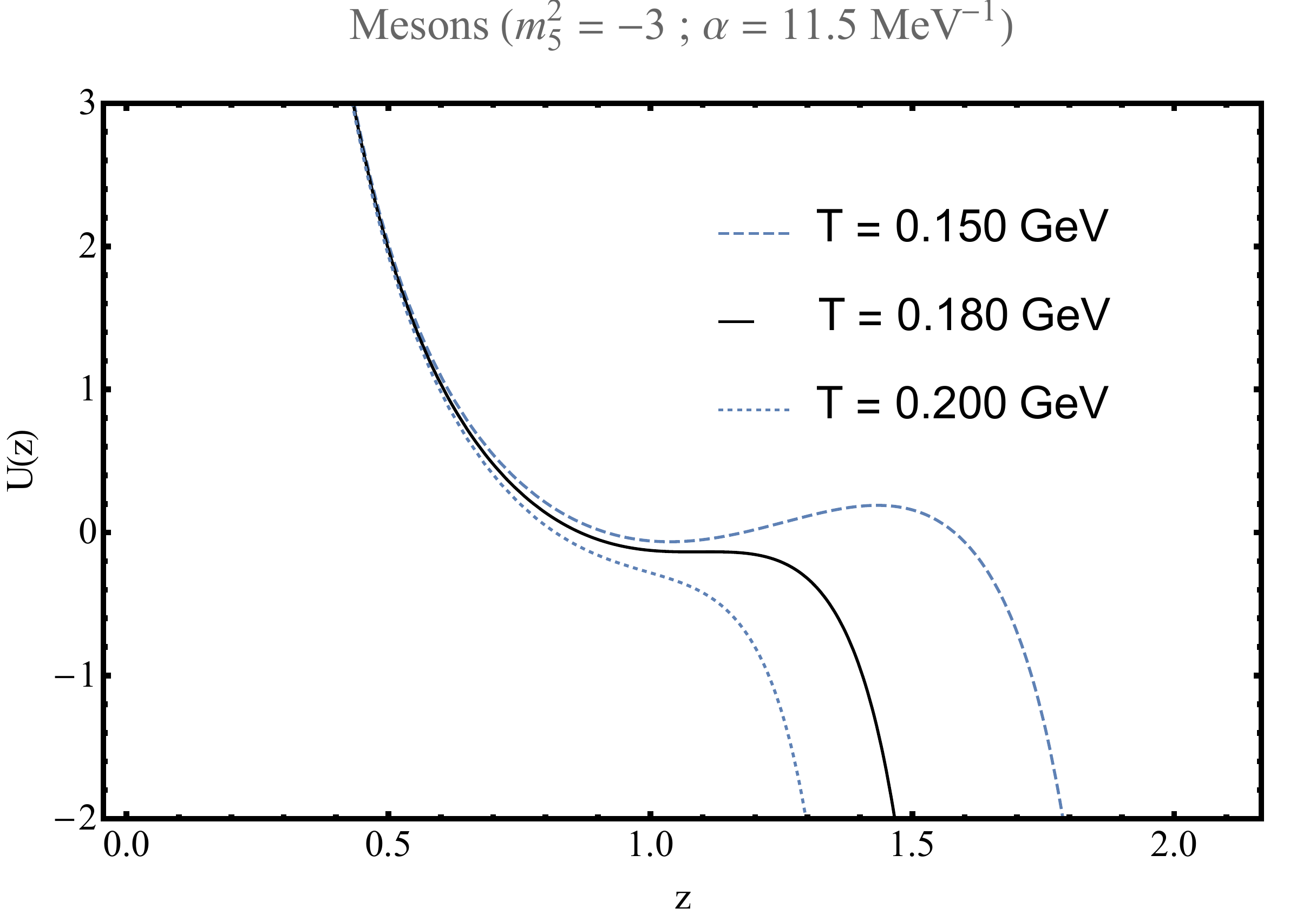}
  \end{tabular}
\caption{
Plots show the holographic potential calculated by using (\ref{potencial}). We consider the thermal version of one of the dilatons discussed in \cite{Gherghetta:2009ac}. On the right side, the traditional case ($\alpha = 0$) is considered, where the continuous line is calculated with the prediction of the model for melting temperature in this model, which is 85 MeV. On the left side, we plot potentials considering a thermal quadratic dilaton with $\alpha = 11.5$ MeV$^{-1}$, which was adjusted in order to obtain 180 MeV for mesonic melting temperature.}
\end{figure*}
\end{center}

\section{Thermal Dilaton}

As in several AdS/QCD models at zero temperature, here we do not consider dynamic dilatons. But if we do not forget that these $A(z)$ and $f(z)$ functions (which define the background) are related to the dilaton,  we can motivate a temperature dependent behavior for it. This is what we call the thermal dilaton.

In this paper, we consider the thermal version of the usual quadratic dilaton and one of the dilatons studied in \cite{Gherghetta:2009ac}. These two dilatons can be summarized as
\begin{equation}
\label{DilatonTermico1}
\phi_{1}(z,T) = \kappa^{2} (1 + \alpha\, T) z^{2},
\end{equation}
and
\begin{equation}
\label{DilatonTermico2}
\phi_{2}(z,T) = \kappa_{1}^{2} (1 + \alpha\, T) z^{2}~\tanh[\kappa_{2}^{2} (1 + \alpha \,T) z^{2}].
\end{equation}

As can be seen, setting $\alpha=0$ implies that both expressions are reduced to the standard form used by other authors, where $\phi_{1}(z,0)$ is the traditional quadratic one and $\phi_{2}(z,0)$ is one of the dilatons used in \cite{Gherghetta:2009ac}. The last one interpolates between two quadratic forms to incorporate chiral symmetry breaking in soft wall models.

In both cases, $\alpha$ is a parameter fixed by using the melting temperature related to light mesons, which is close to $180$ MeV. The parameter $\alpha$ is the only one introduced in our extension of static dilatons used at $T=0$. Values used for $\alpha$ are zero and $27$ MeV$^{-1}$ for dilaton 1; and zero and $11.5$ MeV$^{-1}$ for dilaton 2. On the other hand, since at $T=0$ we recover dilatons used by another author, then we consider parameters that they used in their papers. Therefore, in this case, $\kappa = 0.389$ MeV, $\kappa_{1} = 0.390$ MeV and $\kappa_{2} = 0.428$ MeV. 

In addition to the thermal dilatons (\ref{DilatonTermico1}) and (\ref{DilatonTermico2}), we use a black hole metric (with $R=1$) in $AdS_{5}$, i.e.,
\begin{equation}
A\left(z\right) = \ln \left( \frac{1}{z} \right),\,~~ f\left(z\right) = 1 - (\pi\, T)^{4} z^{4}.
\end{equation}

Thus, we now have all the ingredients to use (\ref{potencial}) in order to define a holographic thermal potential for each dilaton and also to calculate the melting temperature.




\section{Results and conclusions}

An equation such as (\ref{EOM}) can be transformed in a Schr\"odinger-type equation using a Liouville transformation \cite{Miranda:2009qp, Bellantuono:2014lra, Bellantuono:2014lra, Vega:2017dbt, Ibanez:2018xci} or by means of a Bogolyubov transform, as was considered by authors in \cite{Fujita:2009ca} and also by us here. In both cases, it is possible to obtain a well in the holographic potential with a depth which is reduced when the temperature is increased. The well disappears at a temperature interpreted as the melting temperature.

The main difference between the potentials obtained by considering the transforms discussed in the previous paragraph is that, in the case in which we use Bogolyubov transforms, it is necessary to know the hadron mass in order to analyze the potential. In \cite{Fujita:2009ca}, authors have considered mass values calculated at zero temperature, and as it was studied in \cite{Vega:2017dbt, Ibanez:2018xci} for the quadratic dilaton case, the values for melting temperature are close to those achieved from the potential obtained by considering the Liouville transform (which is generally not easy to implement) and also are close to the ones calculated directly from the spectral function. Since  the potential (\ref{potencial}) can be obtained easily for different metrics and dilatons, we consider this procedure a good tool to do a first approach or a quick check of AdS/QCD models at finite temperature.

The $\alpha$ parameter is fixed in order to produce a mesonic melting temperature of $180$ MeV, and as this is the only parameter in our extension we can calculate melting temperature for other scalar hadrons. For $\phi_{1}(z,T)$, when we consider $\alpha = 0$ MeV$^{-1}$, the melting temperature is $48$ MeV for mesons, $36$ MeV for glueballs, $31$ MeV for hybrid mesons and $27$ MeV for tetraquarks; while for $\alpha = 27$ MeV$^{-1}$ we have $180$ MeV for mesons, $96$ MeV for glueballs, $65$ MeV for hybrid mesons and $50$ MeV for tetraquarks. On the other hand, for $\phi_{2}(z,T)$ when $\alpha = 0$ MeV$^{-1}$ we have $85$ MeV for mesonic melting temperature, $30$ MeV for glueballs, $27$ MeV for hybrid mesons and $27$ MeV for tetraquarks. In the case with $\alpha = 11.5$ MeV$^{-1}$, the melting temperatures are $180$ MeV for mesons, $46$ MeV for glueballs, $39$ MeV for hybrid mesons and $33$ MeV for tetraquarks.

As it is possible to see when $\alpha = 0$ dilaton $\phi_{2}(z,T)$ gives us a melting temperature higher than in the quadratic dilaton case, and in some sense it is an additional improvement considering that it is associated with a model where chiral symmetry breaking can be incorporated at $T=0$, but this melting temperature is still low, and in both cases it is not possible to increase it without ruining mass results of these models; but doing $\kappa_{i}^{2} (1 + \alpha T)$, we transform the old dilatons into thermal ones which, as we have shown, offers an opportunity to easily increase the meson melting temperature and allows us to calculate melting temperatures for other scalar hadrons.

To close, we would like to emphasize the idea that calculating melting temperatures from the analysis of the thermal holographic potential obtained from the Bogolyubov transform offers a quick first test for AdS/QCD models at finite temperature. Additionally, we show that the temperature-dependent dilaton could be a good alternative to extend soft wall models used at zero temperature. And finally, we mention that these ideas could be incorporated into studying hadrons in dense media.




\vspace{0.2cm}
\noindent
\textbf{Acknowledgments: }  The authors acknowledge the financial support of FONDECYT (Chile) under Grants No. 1180753 (A. V) and No. 3180592 (M. A. M).

\end{document}